\documentclass[twocolumn,a4paper,showpacs,preprintnumbers,amsmath,amssymb]{revtex4-1}

\pdfoutput=1
\usepackage{graphicx}
\usepackage{dcolumn}
\usepackage{bm}
\usepackage{times}

\begin{document}


\title{Probing two topological surface bands of Sb$_2$Te$_3$ by spin-polarized photoemission spectroscopy}
\author{C. Pauly$^1$}
\email[] {pauly@physik.rwth-aachen.de}
\author{G. Bihlmayer$^2$}
\author{M. Liebmann$^1$}
\author{M. Grob$^1$}
\author{A. Georgi$^1$}
\author{D. Subramaniam$^1$}
\author{M. R. Scholz$^3$}
\author{J. S\'anchez-Barriga$^3$}
\author{A. Varykhalov$^3$}
\author{S. Bl\"ugel$^2$}
\author{O. Rader$^3$}
\author{M. Morgenstern$^1$}
\affiliation{ $^1$II. Physikalisches Institut B and JARA-FIT, RWTH Aachen University, D-52074 Aachen, Germany\\
$^2$Peter Gr\"{u}nberg Institut (PGI-1) and Institute for Advanced Simulation (IAS-1), Forschungszentrum J\"{u}lich and JARA, D-52425 J\"{u}lich, Germany\\
$^3$Helmholtz-Zentrum Berlin f\"ur Materialien und Energie, Elektronenspeicherring BESSY II, Albert-Einstein-Str. 15, D-12489 Berlin, Germany}

\date{\today}

\begin{abstract}
Using high resolution spin- and angle-resolved photoemission spectroscopy, we map
the electronic structure and spin texture of the surface states of the topological
insulator Sb$_2$Te$_3$. In combination with density functional calculations (DFT), we
directly show that Sb$_2$Te$_3$ exhibits a partially occupied, single spin-Dirac cone 
around the Fermi energy $E_{\rm F}$, which is topologically protected.
DFT obtains a spin polarization of the occupied Dirac cone states of 80-90 \%, which is in 
reasonable agreement with the experimental data after careful background subtraction.
Furthermore, we observe a strongly spin-orbit split surface band at lower energy. This state 
is found at $E-E_{\rm F}\simeq -0.8$\,eV at the $\overline{\Gamma}$-point, disperses upwards, 
and disappears at about $E-E_{\rm F}=-0.4$\,eV into two different bulk bands. Along the 
$\overline{\Gamma}-\overline{\mathrm K}$ direction, the band is located within a spin-orbit 
gap. According to an argument given by Pendry and Gurman in 1975, such a gap must contain a
surface state, if it is located away from the high symmetry points of the Brillouin zone. Thus, 
the novel spin-split state is protected by symmetry, too.
\end{abstract}

\maketitle

\section{Introduction}
Topological insulators (TIs) are a new phase of quantum matter giving rise to, e.g.,
a quantum spin Hall phase without external magnetic field~\cite{Bernevig,Bernevig2,Konig}.
Large spin orbit (SO) interaction and inversion symmetry lead to nontrivial edge or
surface states which reside in a bulk energy gap and are protected by time reversal symmetry.
This new phase is classified by a Z$_2$ topological invariant, which distinguishes it
from an ordinary insulator~\cite{Kane}. In three dimensions, the surface states form
an odd number of massless Dirac cones exhibiting a chiral relationship between spin
and momentum of the electrons~\cite{Fu,Moore,Roy,Fu2,Murakami}. Bi$_{1- \rm x}$Sb$_{\rm x}$ was
the first 3D TI to be theoretically predicted~\cite{Fu,Moore} and experimentally
discovered~\cite{Hsieh, Hsieh2}. Subsequent calculations showed that the thermoelectric
materials Bi$_2$Se$_3$, Bi$_2$Te$_3$ and Sb$_2$Te$_3$ should exhibit even simpler TI
properties with only one Dirac cone around $\overline{\Gamma}$~\cite{Zhang}.
Afterwards, Bi$_2$Se$_3$ and Bi$_2$Te$_3$ were studied by spin- and angle-resolved
photoemission spectroscopy (spinARPES) confirming the TI properties of the surface
state~\cite{Xia,Chen,Hsieh3,Hsieh4,Scholz,Pan}.
Scanning tunneling spectroscopy (STS) revealed the absence of backscattering, i.e.\
momentum inversion, for the surface state of these materials, which confirms the
protective spin chirality  of the Dirac cone~\cite{Roushan,Zhang2,Alpichshev}.
While the TI nature of Bi$_2$Se$_3$ and Bi$_2$Te$_3$ is well established, the phase
change material Sb$_2$Te$_3$~\cite{Wuttig} was rarely probed. Recent ARPES
measurements suggest that the Fermi level, $E_{\rm F}$, of single-crystal Sb$_2$Te$_3$
is within the bulk valence band probably due to hole doping~\cite{Hsieh4}.
In contrast, thin films grown by molecular beam epitaxy (MBE) exhibited a Dirac
cone around $E_{\rm F}$ in ARPES \cite{Wang}. However, the spin chirality and 
the topological nature of the Dirac cone has not been tackled so far.

Here, we report spin resolved ARPES data in combination with densitiy functional theory (DFT) 
calculations on single crystal Sb$_2$Te$_3$ showing the spin-polarized nature of the Dirac cone. 
In addition, we found another surface state at lower energy exhibiting strong Rashba-type 
spin-splitting. This splitting described by the Rashba parameter $\alpha \simeq 1.4$ eV\AA\;is larger 
than for Au(111)~\cite{LaShell,Hoesch} or Bi(111) \cite{Ast,Koroteev,Kimura}, but lower 
than in Bi-based surface alloys~\cite{Ast1,Ast2}. DFT calculations reproduce this state and reveal 
that it is located in a narrow gap in the $\overline{\Gamma}-\overline{\mathrm K}$ direction that 
originates from SO interaction. This observation is in line with the analytical prediction from 1975~\cite{Pendry},
that in the case of a SO gap away from high symmetry points of the Brillouin zone, there must be, at least, 
one surface state present. This state is, thus, the second surface state of Sb$_2$Te$_3$, which is protected by 
symmetry.

\section{Experiment}
\begin{figure}[tb]
\includegraphics[width=8.5cm]{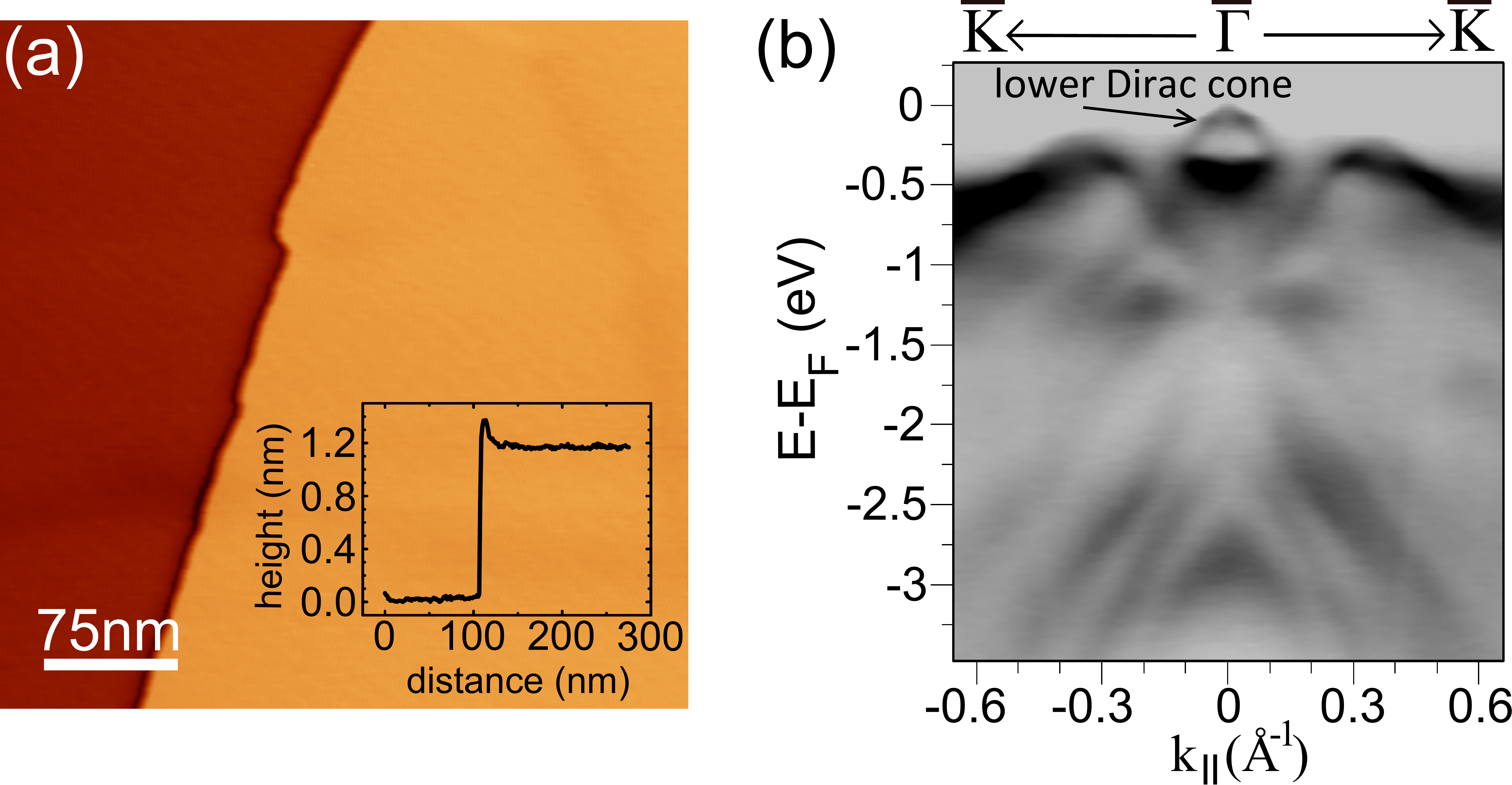}
\caption{\label{figure1}(color online) (a) STM image of cleaved Sb$_2$Te$_3$(0001);
 $V=-1$\,V, $I=0.8$\,nA, $375\times375$\,nm$^2$. Inset: Line profile showing the step which corresponds to the height of one QL. (b) ARPES data of
 Sb$_2$Te$_3$(0001) along $\overline{\Gamma}-\overline{\mathrm K}$ at an incident
 photon energy $h\nu= 50$\,eV; Dirac cone is marked.}
\end{figure}

Spin- and angle-resolved photoemission (spinARPES) experiments have been performed at 300\,K
with electron analyzers Scienta R8000 and SPECS PHOIBOS 150 using linearly polarized
synchrotron radiation from the beamlines UE112-PGM-1 and UE112-lowE-PGM2 at BESSY II. Energy and momentum re\-so\-lu\-tion of the Scienta R8000 analyzer are 20\,meV and 0.6\,\% of the surface Brillouin zone, respectively. Spin resolution is achieved with a Rice University Mott polarimeter operated at 26\,kV (Sherman function $S_{\rm eff} = 0.16$)
and capable of recording the two orthogonal spin directions in the surface plane of
the sample. The energy and momentum re\-so\-lu\-tion of the spin resolving apparatus are 100\,meV and 4\,\% of the surface Brillouin zone, respectively. Scanning tunneling microscopy (STM) is performed with a modified Omicron STM at 300\,K. For all measurements, the Sb$_2$Te$_3$ single crystal is cleaved in ultra high vacuum (UHV) at a base pressure of $1\cdot 10^{-10}$\,mbar.
Large terraces with widths of several 100 nm separated by step edges of 1 nm height result as verified by STM (Fig.~\ref{figure1}(a)). Figure~\ref{figure1}(b) reveals that ARPES measurements exhibit distinct bands after cleavage, however with a relatively strong dependance on photon energy (see below). 

The calculations are performed within the generalized gradient approximation~\cite{PBE}
to DFT, employing the full-potential linearized augmented planewave method as implemented
in the {\sc Fleur} code~\cite{Fleur}. SO coupling is included in a non-perturbative
manner~\cite{Li}. Based on the optimized bulk lattice parameters, the surfaces are
simulated by films of a thickness of six quintuple layers (QLs) embedded in vacuum.

\section{General characterization}
\begin{figure}[b]
\includegraphics[width=8.6cm]{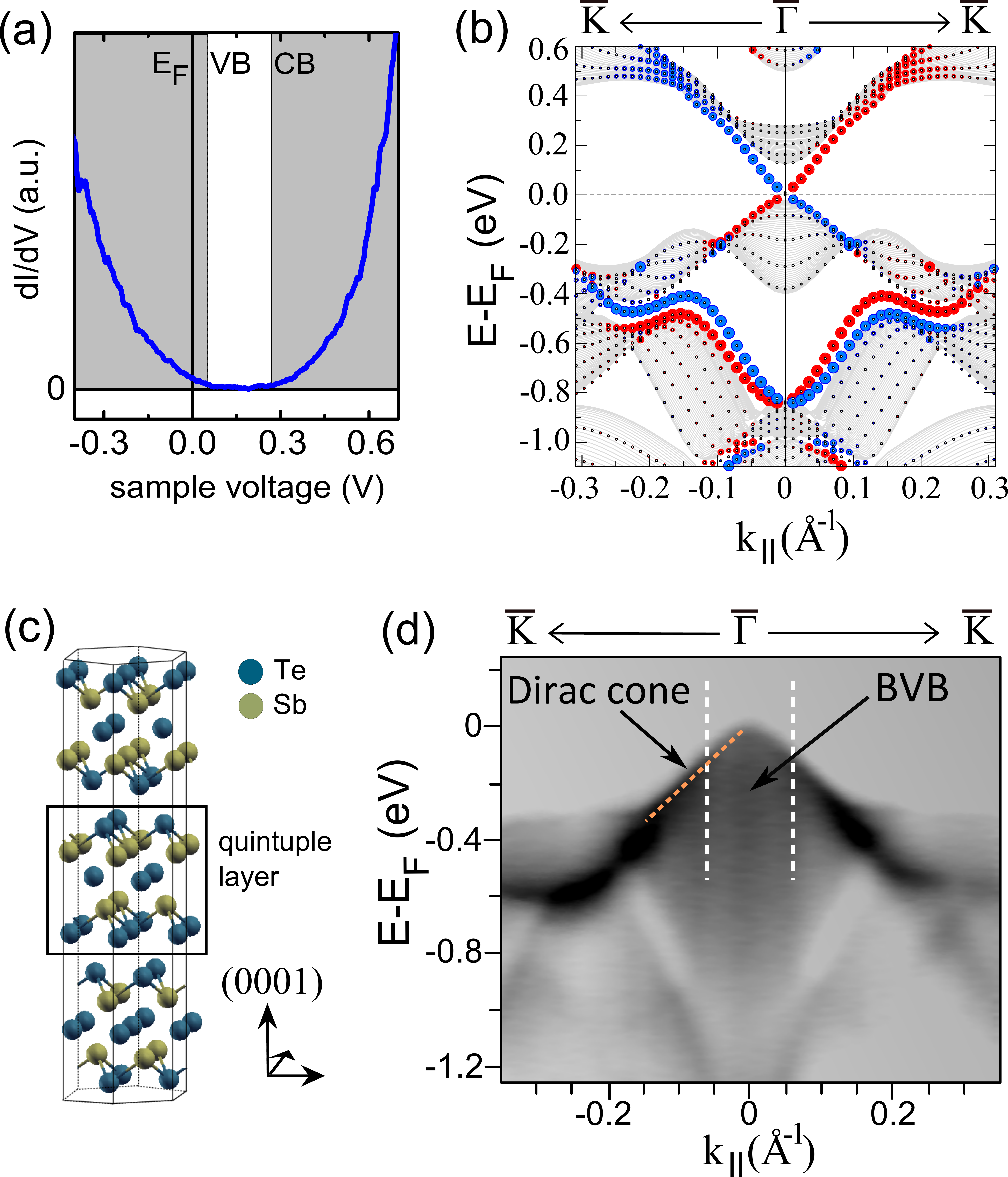}
\caption{\label{figure2}(color online) (a) $dI/dV$ spectrum recorded by STM on the cleaved Sb$_2$Te$_3$(0001) surface;
 $V_{\rm stab}=-0.5$\,V, $I_{\rm stab}=1$\,nA, $V_{\rm mod}=10$\,mV; band edges of the
 valence band (VB) and conduction band (CB) are marked by lines, Fermi level $E_{\rm F}$ is marked. (b) Band structure in
 $\overline{\Gamma}-\overline{\mathrm K}$ direction as calculated by DFT including spin-orbit coupling; states resulting from a film calculation
 are shown as circles with the color (blue or red) indicating different spin directions
 and the size of colored circles marking the magnitude of the spin density (for absolute spin polarization values see Fig. \ref{figure4}); shaded
 areas are projected bulk bands originating from a bulk calculation. (c) Sketch of
 the crystal structure of Sb$_2$Te$_3$; one QL is marked with different atoms
 in different colors as indicated. (d) ARPES data of the lower Dirac cone along $\overline{\Gamma}-\overline{\mathrm K}$ and bulk
 valence band (BVB) as marked; $h\nu=55$\,eV; orange, dashed line is a guide to the eye from which the Fermi velocity $v_{\rm F}$ is deduced; white dashed lines mark positions of the energy distribution curves (EDCs) shown in Fig. \ref{figure3}(a), (b) and (c).}
\end{figure}

The sketch in Fig.~\ref{figure2}(c) shows the crystal structure of Sb$_2$Te$_3$ consisting of consecutive QLs with stacking
sequence Te(1)-Sb-Te(2)-Sb-Te(1). The different numbers in parentheses mark the different
environments of the Te layers. The coupling within a QL is strong, whereas the interaction
between two QLs is predominantly of van der Waals type~\cite{Zhang}. Consequently,
cleavage leads to a Te terminated (0001) surface with hexagonal symmetry as has been
verified by low-energy electron diffraction. Identically to Bi$_2$Se$_3$ and Bi$_2$Te$_3$,
Sb$_2$Te$_3$ has inversion symmetry with the layer Te(2) containing the center of inversion.
This simplifies the calculation of Z$_2$, which becomes an analysis of states at the high
symmetry points only, and, thus, leads to the straightforward identification of a strong
topological insulator~\cite{Fu2,Zhang}.

The electronic structure of Sb$_2$Te$_3$ is firstly probed by STS as shown in Fig.~\ref{figure2}(a).
STS records the differential tunneling conductivity $dI/dV$ which is proportional to
the local density of states (LDOS) of the sample~\cite{Morgenstern}. As theoretically
predicted~\cite{Zhang}, a small band gap of about 200\,meV is observed. The Fermi level $E_{\rm F}$
(sample bias $V=0$\,V) is found close to the valence band edge indicating
p-type doping \cite{Hsieh4,Wang} in agreement with a recent STM study on MBE grown thin films \cite{Jiang}. The band structure from DFT calculations of a six QL film including SO
interaction is shown in Fig.~\ref{figure2}(b). It is in good agreement with
a recent calculation~\cite{Eremeev}. Small differences to earlier calculations~\cite{Zhang,Hsieh4}
can be traced back to the sensitivity of the electronic structure to small changes of the geometrical parameters.
The spin-polarized states are displayed as colored circles being blue or red for the
different spin orientations. Only the spin polarization perpendicular to the in-plane wave vector of the electrons $\mathbf k_\|$ and
the surface normal is shown. The varying radius marks the absolute value of this k-resolved spin density at and above the surface. The absolute spin polarization of the states in comparison with the values from the experiment is discussed in more detail in Fig.~\ref{figure4}. Projected bulk bands resulting from a bulk
calculation are shown as shaded areas Fig.~\ref{figure2}(b). A single Dirac cone originating from topological protection is found around $\overline{\Gamma}$
with Dirac point at $E_{\rm F}$ and overlap of the occupied states with bulk states. Furthermore, we checked the Z$_2$ invariant with the help of our DFT calculations and found that Sb$_2$Te$_3$ is indeed topologically non-trivial with an index of (1;000).
Another bulk band gap exists around $E-E_{\rm F}=- 400$\,meV. It houses two spin-polarized surface
states exhibiting a Rashba-type spin splitting $\Delta E = \alpha \cdot |k_\||$ with the wave number parallel to the surface $k_{\|}$ and the Rashba coefficient $\alpha \simeq 1.4$ eV\AA, at least, up to $k_\|\simeq 0.05$ ${\rm{\AA^{-1}}}$. This $\alpha$ is larger  than the
value for Au(111) ($\alpha = 0.33$ eV\AA)~\cite{LaShell} or Bi(111) ($\alpha = 0.55$ eV\AA)~\cite{Koroteev}, both consisting of heavier atoms, but lower
than the largest $\alpha$-values so far found in Bi surface alloys ($\alpha = 3.8$ eV\AA)~\cite{Ast1}.

\section{Dirac cone in fundamental band gap}
\begin{figure}[b]
\includegraphics[width=8.6cm]{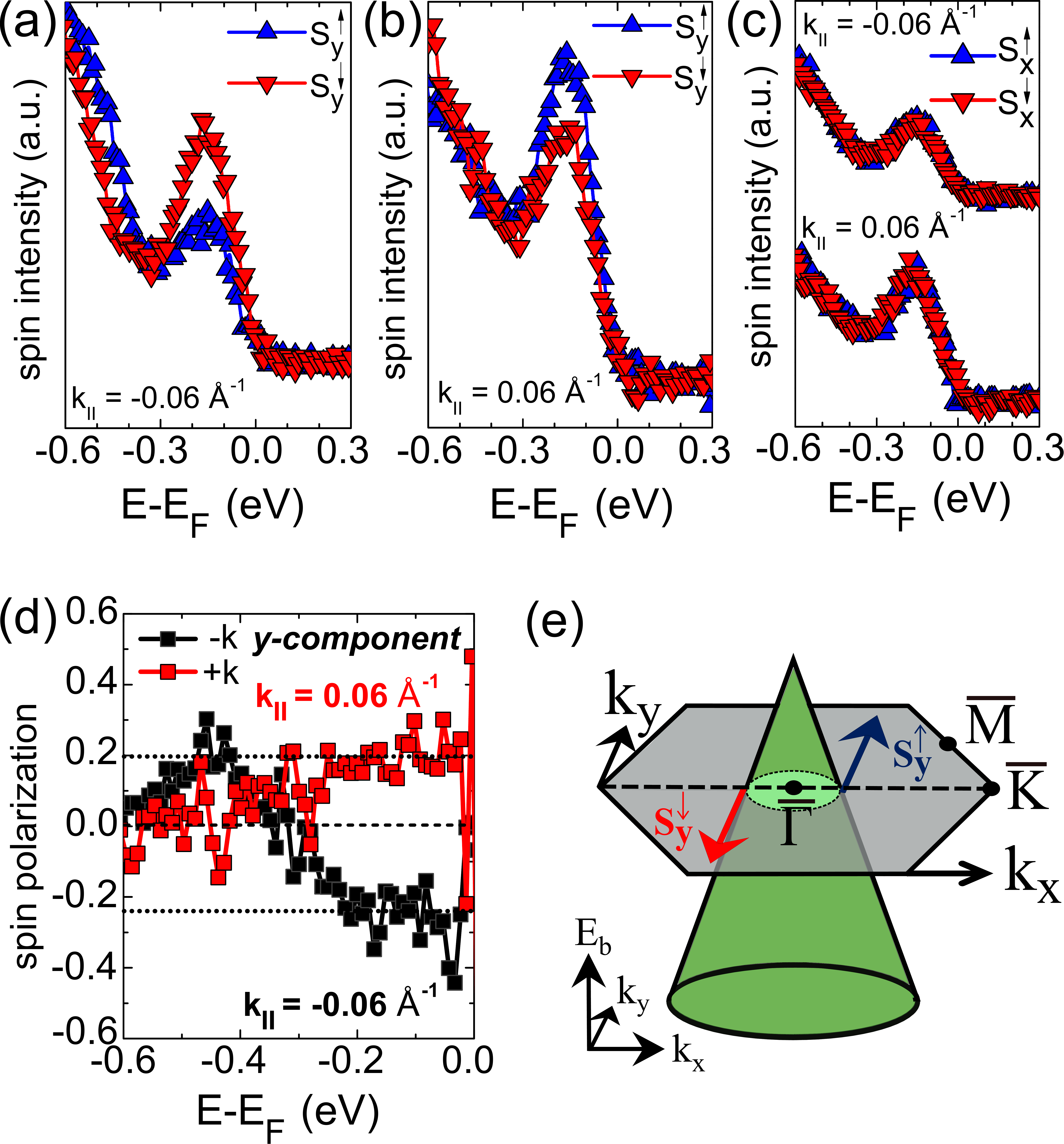}
\caption{\label{figure3}(color online) (a), (b) Spin-resolved energy distribution curves (EDCs) for the spin component perpendicular to $\mathbf k_{\rm \|}$
recorded at $k_{\rm \|}$-values as indicated and marked by dashed lines in Fig.~\ref{figure2}(d); different colors mark different spin directions; $h\nu = 54.5$ eV.
(c) Spin-resolved EDCs for the spin component parallel to $\mathbf k_{\rm \|}$;
$h\nu = 54.5$ eV. (d) Resulting spin polarization perpendicular to the two different $k_{\rm \|}$ as marked.
(e) Sketch of the lower Dirac cone with spin directions marked as deduced from spinARPES and in accordance with DFT.}
\end{figure}
Figure~\ref{figure2}(d) shows the measured energy dispersion around the Dirac point
along $\overline{\Gamma}-\overline{\mathrm K}$, which is most easily visible at $h\nu \approx 55$ eV. The linear dispersion of the Dirac cone with a
crossing point close to $E_{\rm F}$ is visible. The latter indicates that the position of the surface Fermi level
is predominantly determined by the Dirac electrons and not by extrinsic doping.\;This deviates from previous ARPES results obtained on bulk Sb$_2$Te$_3$ \cite{Hsieh4}, but is
in agreement with ARPES data from thin films  grown by MBE \cite{Wang}. It points to a low defect
density of the investigated crystal \cite{Jiang}. The linear dispersion is fitted by $E-E_{\rm F}= \hbar v_{\rm F} |k_\||$
resulting in Fermi velocity $v_{\rm F}=3.8\pm 0.2 \cdot 10^{5}$\,m/s (dashed, orange line), which agrees reasonably with $v_{\rm F}= 3.2 \cdot 10^{5}$\,m/s obtained by DFT [Fig.~\ref{figure2}(b)]. The background of bulk valence bands (BVB) found in DFT is also visible in Fig.~\ref{figure2}(d).
In order to verify the chiral spin polarization of the Dirac cone, we use spinARPES sensitive to
the spin components within the surface plane.

\begin{figure}[tb]
\includegraphics[angle=0,width=8cm]{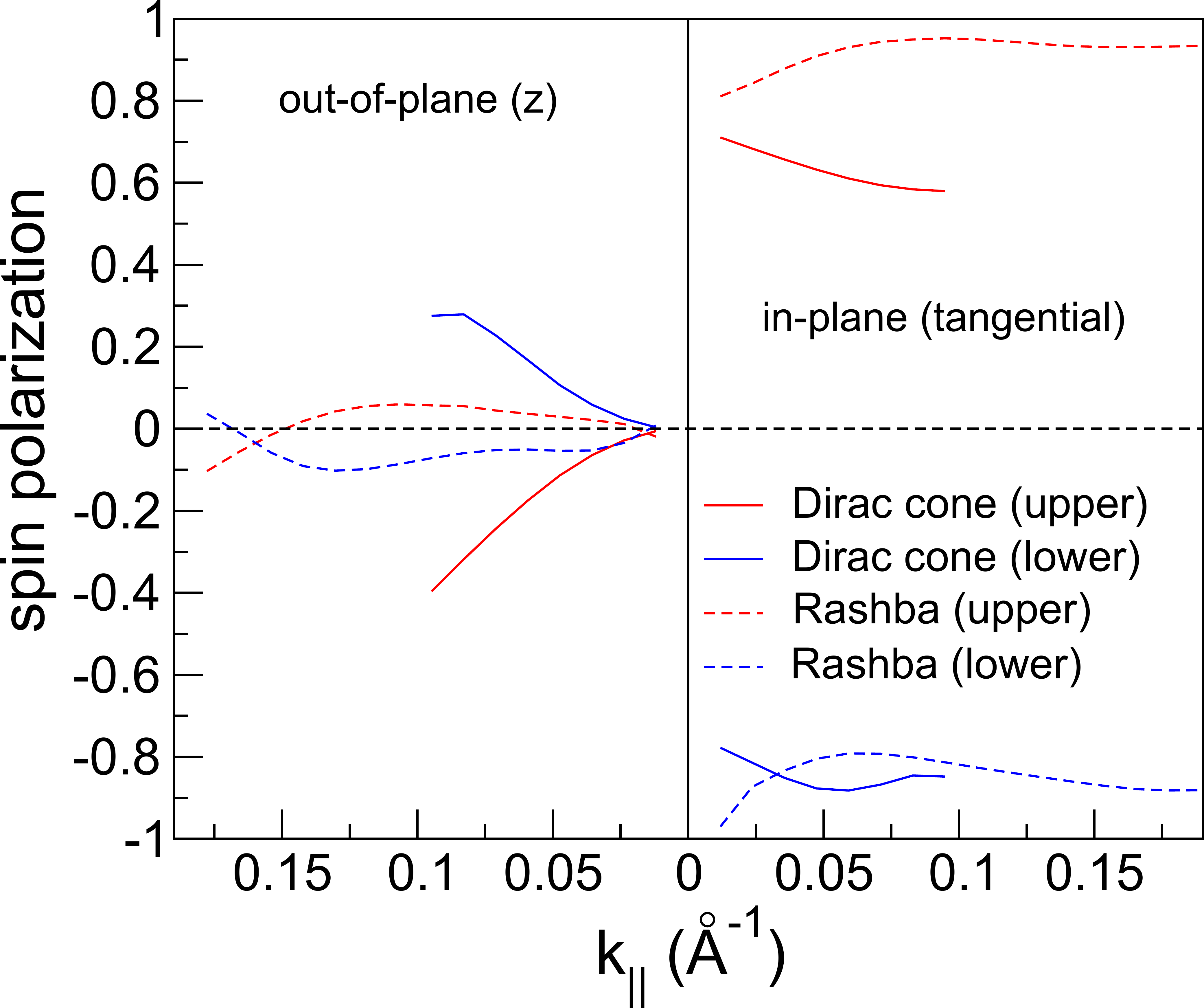}
\caption{\label{figure4}(color online) Expectation values from DFT for the out-of-plane and in-plane component of the spin polarization along
$\overline{\Gamma}-\overline{\mathrm K}$ for the Dirac cone and the Rashba-type surface state (Rashba). Upper and lower indicates the energies above and below the Dirac point and the energetically higher and lower Rashba band, respectively.}
\end{figure}
Figure~\ref{figure3}(a), (b) and (c) show the spin-resolved energy distribution curves (EDCs) as
measured at $k_{\rm \|}=-0.06$\,$\rm{\AA^{-1}}$ and $k_{\rm \|}=0.06$\,$\rm{\AA^{-1}}$ as marked.
The spin component perpendicular to $\mathbf{k}_{\rm \|}$ and the surface normal [Fig.~\ref{figure3}(a),
(b)] exhibits an intensity difference between spin up and spin down component which reverses for
the opposite wave number. In contrast, the spin component parallel to $\mathbf{k}_{\rm \|}$
[$S_x$, Fig.~\ref{figure3}(c)] shows no spin polarization.
This leads to the spin momentum relation depicted in Fig. \ref{figure3}(e) for the in-plane spin polarization which is typical for a topological insulator. The
spin is perpendicular to $\mathbf{k}_{\rm \|}$ and rotates counterclockwise for the lower part of
the Dirac cone as in the case of Bi$_2$Te$_3$, Bi$_2$Se$_3$ \cite{Xia, Chen, Hsieh3, Hsieh4} or the tunable topological insulator BiTl(S$_{1- \mathrm{\delta}}$Se$_{\mathrm \delta}$)$_2$ \cite{Xu}. The same sense of rotation is found by the DFT calculations [Fig.~\ref{figure2}(b)].
Figure~\ref{figure3}(d) shows the resulting spin polarization for the two opposite momenta calculated
according to $P_y=(S^{\uparrow}_{y}-S^{\downarrow}_{y})/(S^{\uparrow}_{y}+S^{\downarrow}_{y})$
with $S^{\uparrow}_{y}$ and $S^{\downarrow}_{y}$ being the measured spin resolved intensities perpendicular
to $\mathbf{k}_{\rm \|}$. The polarization of $P_y\simeq 20$ \% has opposite sign for opposite momenta.

The spin polarization of the Dirac cone in Sb$_2$Te$_3$ is further calculated by DFT and analyzed in terms of
an in-plane and an out-of-plane component. 
\begin{figure*}[tb]
\includegraphics[width=17.2cm]{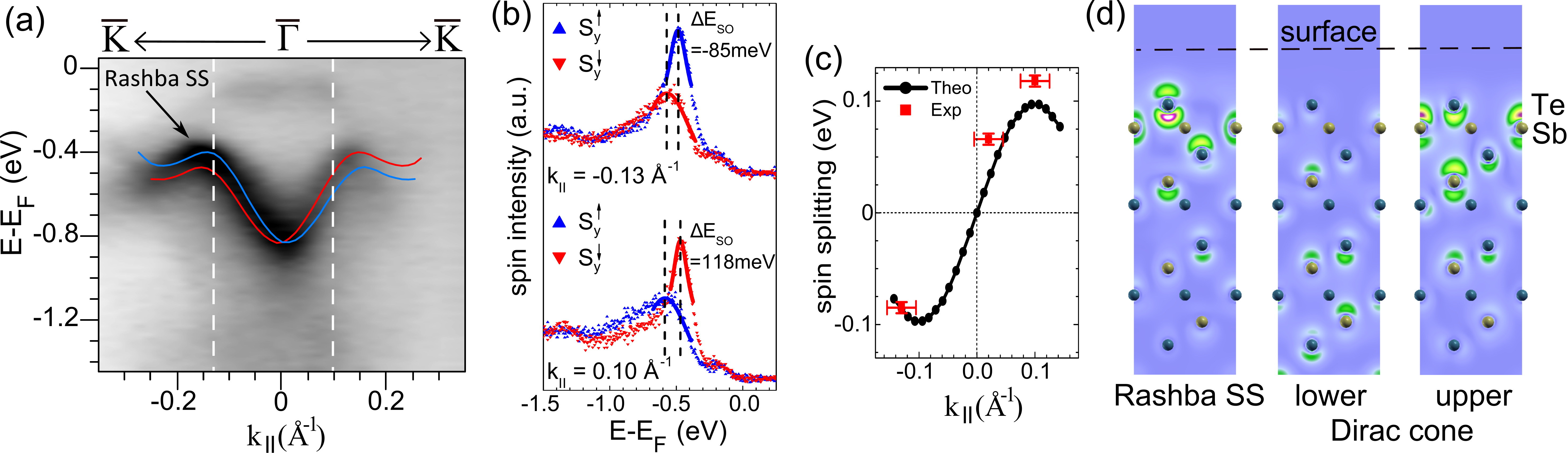}
\caption{\label{figure5}(color online) (a) ARPES dispersion of the Rashba type surface state (Rashba SS) along
$\overline{\Gamma}-\overline{\mathrm K}$ direction; $h\nu=22$\,eV; band structure of the Rashba surface states from DFT is superimposed as blue and red lines; dashed lines mark the position of the EDCs in (b). (b) Spin resolved EDCs (points in red and blue for the two spin directions perpendicular to $\mathbf{k}_{\rm \|}$) at different momenta as indicated and marked by dashed lines in (a); fits of the peaks are shown as solid lines (red, blue); peak positions are marked by dashed lines and
lead to the spin splitting energies $\Delta E_{\rm SO}$ indicated; (c) calculated spin splitting (Theo) of the Rashba state in comparison with measured spin splitting (Exp).
(d) Two-dimensional cut through the calculated local density of states for the Rashba state and the lower and upper part of the Dirac cone at $k_{\rm \|}=0.06$\,$\rm{\AA^{-1}}$.}
\end{figure*}
Only the in-plane component perpendicular to $\mathbf{k}_{\rm \|}$ is considered,
since we did not find any spin polarization for the direction parallel to $\mathbf{k}_{\rm \|}$. Figure \ref{figure4} shows the resulting spin polarization values with respect to the wavenumber integrated over the first two atomic layers. This area approximately corresponds to the penetration depth in the ARPES
experiment at a photon energy of 55\,eV. While $P_y \simeq 1$ is found for free Dirac cones by DFT \cite{Pan}, we find a reduced polarization for the lower Dirac cone of roughly 80\,\% near the $\overline{\Gamma}$ point, which increases to about 90\,\% at $k_{\rm \|}=0.06$\,$\rm{\AA^{-1}}$. This is mostly due to a penetration of the Dirac cone states into subsurface layers (Fig. \ref{figure5}(d)) where fluctuating electric fields lead to a complex spin texture. In contrast, the in-plane
polarization of the upper Dirac cone decreases towards higher wavenumbers down to 60\,\%. Note, the considerable out-of-plane polarization in $\overline{\Gamma}-\overline{\mathrm K}$ direction for higher wavenumbers, which is in line with the warping of the Dirac cone at higher energies \cite{Chen, Kong}. This result agrees also qualitatively with the calculations of Yazyev et al. \cite{Yazyev} for Bi$_2$Te$_3$ and Bi$_2$Se$_3$. However, the polarization values in those materials are typically smaller, reflecting the stronger spin-orbit entanglement caused by the heavier Bi atom. 

The discrepancy in the in-plane spin polarization between calculation ($P_y \simeq 90$ \%) and experiment ($P_y \simeq 20$ \%) at $k_{\rm \|}=0.06$\,$\rm{\AA^{-1}}$ can be traced back to the finite angular resolution of the spinARPES experiment. Deconvolution from the BVB reveals an estimated spin polarization for the lower Dirac cone of 80-90 \% in experiment (see appendix), which is in a good agreement with the DFT result. 

\section{Rashba spin-split surface state protected by spin-orbit}
Next, we demonstrate that Sb$_2$Te$_3$ exhibits an additional spin split surface state originating
from SO interactions. Figure~\ref{figure5}(a) shows ARPES data along $\overline{\Gamma}-\overline{\mathrm K}$
recorded at lower photon energy, $h\nu=22$\,eV, and superimposed by the DFT band structure of the
Rashba-type band. Based on the excellent concurrence, we conclude that the ARPES data at $h\nu=22$ eV are dominated by these
bands, while the Dirac cone is barely visible. We determine the spin splitting of the Rashba bands using spin-resolved EDCs for the spin direction perpendicular
to $\mathbf{k}_{\rm \|}$ as shown in Fig.~\ref{figure5}(b). Therefore, we fit the peak in
each curve by a Lorentzian function as shown by solid lines. This leads to the spin splitting
energies $\Delta E_{\rm SO}$ indicated and plotted for different $\mathbf{k}_{\rm \|}$ in Fig. \ref{figure5}(c).
Reasonable agreement between theoretical and experimental $\Delta E_{\rm SO}$ is found. Moreover,
as expected for a Rashba-type spin splitting, the spin direction for upper and lower peak inverts
by inverting the $\mathbf{k}_{\rm \|}$ direction. We checked that negligible spin polarization
is found parallel to $\mathbf{k}_{\rm \|}$ which implies that the spin of the upper (lower) band
rotates clockwise (counter-clockwise) with respect to $\mathbf{k}_{\rm \|}$ in agreement with the
DFT results. As depicted in Fig.~\ref{figure4}, DFT further reveals that the Rashba state shows no pronounced out-of-plane polarization within the first two atomic layers whereas in the in-plane direction, the different spin branches are nearly fully spin polarized. From $k_{\rm \|}=0.05$\,$\rm{\AA^{-1}}$ to $k_{\rm \|}=0.15$\,$\rm{\AA^{-1}}$ the lower spin branch shows a slightly lower in-plane polarization than the upper spin branch. This is probably due to the proximity to the bulk band. Altough there is a small reduction of the polarization, the coupling to the bulk band seems to be rather low. Otherwise, the reduction should be more pronounced. Compared to the DFT, the experimental spin resolved data reveal a net spin polarization of $P_y\simeq 45$ \% for the upper band and $P_y\simeq 18$ \% for the lower band (Fig.~\ref{figure5} (b)). Most likely, the overlap of the two bands in the experiment as well as its overlap with the bulk bands caused by the limited energy and momentum resolution of the spin detector is responsible for the small numbers. The fact that the peak at higher energy is sharper is probably related to its larger separation from the bulk bands [as visible in Fig.~\ref{figure2}(b)] leading to longer lifetime.

\begin{figure*}[tb]
\includegraphics[width=16cm]{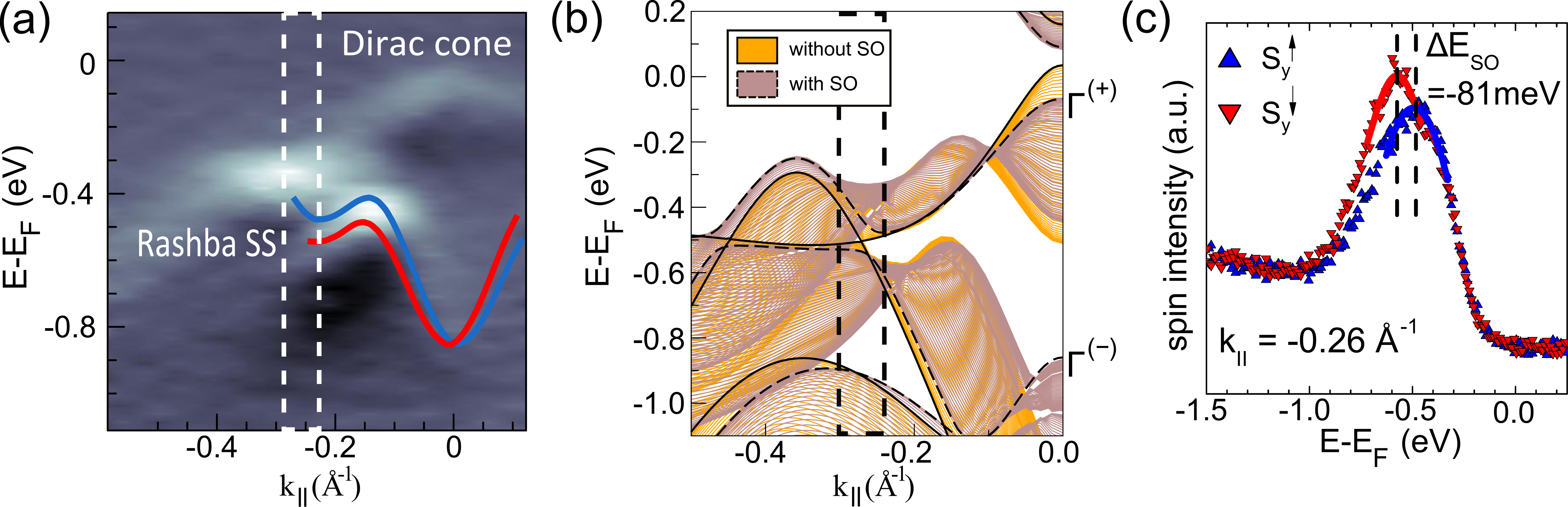}
\caption{\label{figure6}(color online) (a) ARPES data including the Rashba state at higher $|\mathbf{k}_{\rm \|}|$; $h\nu= 54.5$ eV; for better visibility, the derivative with respect to $E$ is shown; Dirac cone and Rashba-type surface state (Rashba SS) with superimposed  $E(\mathbf{k}_{\rm \|})$ curves of the Rashba states (colored lines) from DFT calculation are marked. Dashed boxes indicate the same area in (a) and (b) as well as the $|\mathbf{k}_{\rm \|}|$-position of the EDC in (c). The width is given by the angular resolution of the experiment. (b) Bulk band structure along $\overline{\Gamma}-\overline{\mathrm K}$ calculated with (gray lines) and without (yellow lines) SO interaction; black lines mark the band edges with (dashed) and without (solid) SO interaction; with SO interaction, a gap opens near the center of the graph around $E=-0.5$\,eV and
$k_{\rm\|}=\pm0.26$\,$\rm{\AA^{-1}}$; $\Gamma^{(-)}$, $\Gamma^{(+)}$ mark the parity of the two bands at the $\Gamma$ point;
(c) Spin resolved EDCs measured at the position of the SO gap ($|\mathbf{k}_{\rm \|}|$ as marked); $h\nu =54.5$\,eV; peak positions
as determined from Lorentzian fits (solid lines) are indicated by dashed lines; the resulting $\Delta E_{\rm SO}$ is marked.}
\end{figure*}
The charge density of the Rashba-type state, as shown for $k_{\rm \|}=0.06$\,$\rm{\AA^{-1}}$ in
Fig.~\ref{figure5}(d) reveals that it has predominantly Te $p_z$ character and is localized strongly
within the Te surface layer. In contrast, the states of the Dirac cone are more Sb $p_z$ like and
are penetrating more strongly into the bulk of Sb$_2$Te$_3$.
The different penetration might be the reason why the Dirac cone is more easily observed at higher photon energy,
while the Rashba state dominates the spectra at $h\nu=22$ eV. 
The electric field between the surface Te-layer and the subsurface Sb-layer is deduced from the
calculated surface core level shift to be about $2 \cdot 10^8$ V/m
having a strong dipolar contribution between Te$^{\delta -}$ and Sb$^{\delta +}$. This surface
dipole is probably responsible for the relatively large Rashba coefficient, similar to the
findings in surface alloys \cite{Ast1} and layered bulk compounds \cite{Ishizaka}.

Differently from the Rashba bands found so far \cite{LaShell, Hoesch}, the DFT calculations shown
in Fig.~\ref{figure2}(b) predict that the different spin branches disperse into different projected
bulk continuum bands. Thus, each spin branch connects the upper and the lower bulk band surrounding
the gap by dispersing from $k_{\rm \|}=-0.28$\,$\rm{\AA^{-1}}$ to $k_{\rm \|}=0.28$\,$\rm{\AA^{-1}}$.
At $\overline{\Gamma}$, the surface state is spin degenerate as requested by time reversal symmetry.

Figure~\ref{figure6}(a) shows the measured band structure close to the point where the Rashba bands merge with the bulk bands according to DFT. Indeed, a  band moving upwards and a band moving downwards are discernible up to about $|\mathbf{k}_{\rm \|}|=0.27$ $\rm{\AA^{-1}}$. Fig.~\ref{figure6}(c)
reveals that a spin splitting of about 81\,meV is still visible at $|\mathbf{k}_{\rm \|}|=0.26$ $\rm{\AA^{-1}}$ (in comparison to a spin splitting of 80\,meV in the DFT), i.e., close to the point where merging of surface bands and bulk bands is obtained in the calculation.
The origin of this remarkable behavior is illustrated in Fig.~\ref{figure6}(b),
where the calculated bulk band structure with and without SO interaction is shown.
Obviously the SO interaction opens a gap between the projected bulk states originating from a band
$\Gamma^{(+)}$ near the Fermi level and  a lower-lying $\Gamma^{(-)}$ band, where $(+)$ and $(-)$ marks the parity of the states at $\overline{\Gamma}$.
The gap is found at $k_{\rm\|}=0.26$\,$\rm{\AA^{-1}}$
along the line $\Gamma-\Sigma$ of the bulk band structure ($\overline{\Gamma}-\overline{\mathrm K}$ in terms of surface Brillouin zone).
In such a SO gap, according to a theoretical argument given by Pendry and Gurman \cite{Pendry}, at least one surface state with two spin channels must exist.
The requirement is that the gap is not located at a high symmetry point of the Brillouin zone as in our case. Thus, the observed Rashba split
surface state, which we experimentally resolve within the spin-orbit gap, is protected by this gap. Similarly to the topologically protected Dirac cone,
it connects the lower and the upper bulk bands, which are inverted by SO interaction.
So far, there has been very little experimental proof of such a surface state within a SO gap away from a high symmetry point. Feder and Sturm \cite{Feder} report of a spin-orbit generated gap along the symmetry line $\overline{\Gamma}  \overline{\mathrm H}$ in W(001) by means of tight-binding calculation, in which a surface state is found experimentally \cite{Willis}. Another example is a well definded surface state located in the $\overline{\Gamma}-\overline{\mathrm T}$ direction in Bi(111) \cite{Jezequel, Yaginuma}. First priciple calculations by Gonze et al. \cite{Gonze} resolved a SO gap on this symmetry line which is consistent in energy with the measured surface state.
Differently from these states, our observation reveals a \textit{spin-splitted} surface state with topological character, which, moreover exists in parallel to a surface Dirac cone in the fundamental gap. Thus, it adds a distinct example to Pendry and Gurman's criteria.

\begin{figure}[tb]
\includegraphics[width=7cm]{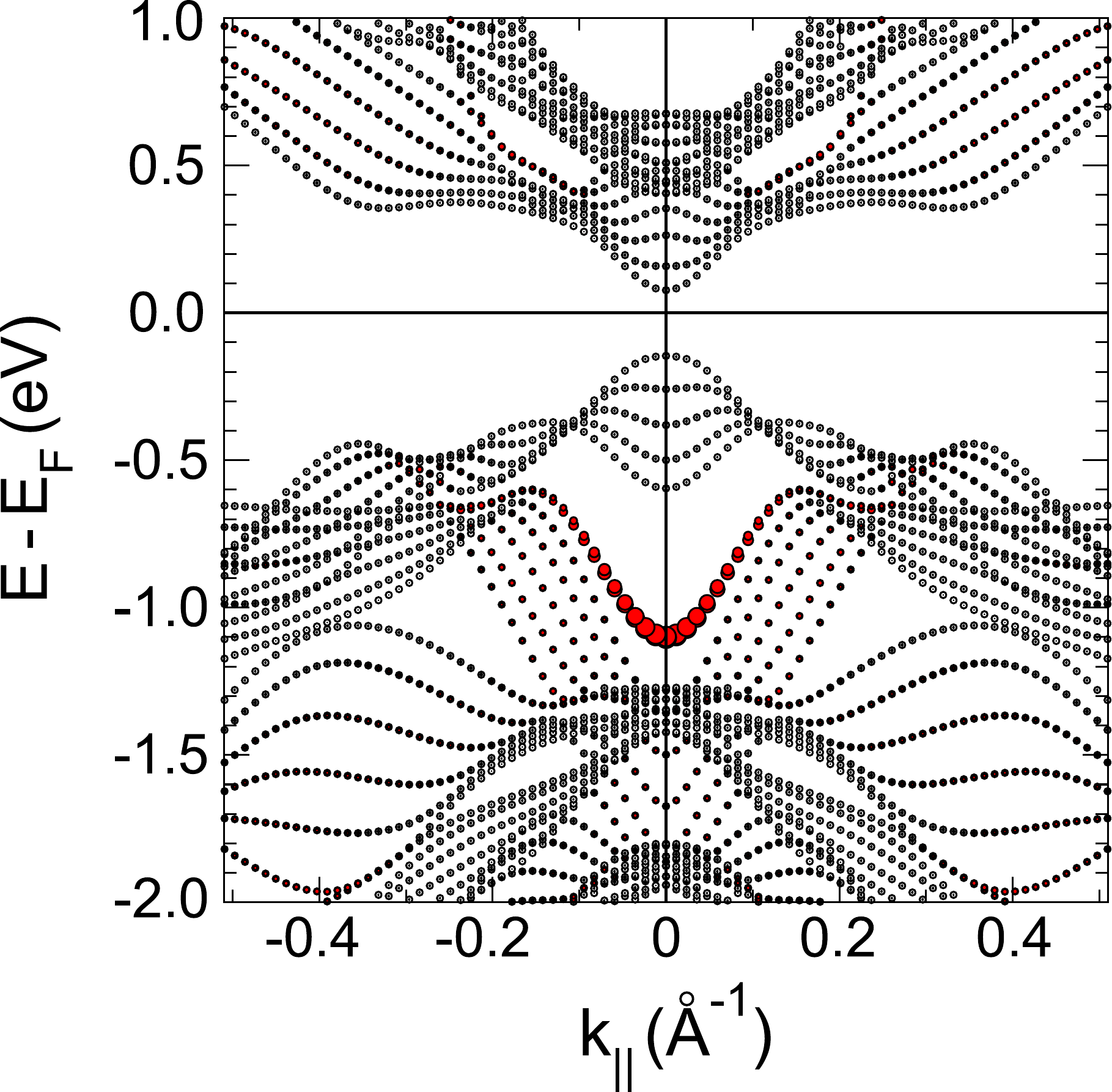}
\caption{\label{figure7}(color online) DFT band structure calculated for a slab without SO coupling. The surface state (red dots) still exists, but is only connected to the lower bulk band exhibiting $\Gamma^{(-)}$ character.}
\end{figure}
Figure~\ref{figure7} shows the band structure of Sb$_2$Te$_3$ from DFT without SO interaction. The surface state at the Fermi level does not exist in this
case, confirming the topological nature of this state. The Rashba state becomes a spin degenerate state and merges for
positive and negative momenta with the same bulk band. Thus, it does not connect the upper and the lower bulk band, i.e., it loses its protected character. This proves that the unconventional behavior towards higher wavenumbers is driven by SO coupling in line with Pendry and Gurman's argument \cite{Pendry}. 

In summary, spinARPES reveals the spin texture of the Dirac cone within the fundamental gap of Sb$_2$Te$_3$,
which rotates counter-clockwise for the lower part of the Dirac cone. A low defect density of the crystal
allows to follow this state up to the Dirac point. In addition, in accordance with DFT
calculations, we identified a novel, strongly spin-split Rashba-type surface state which is protected by a
spin-orbit gap away from $\overline{\Gamma}$ and connects an upper and a lower bulk valence band. This state is protected by symmetry according to a fundamental criterion given by Pendry and Gurman in 1975.

\section{acknowledgment}
We gratefully acknowledge provision of the sample by M. Wuttig and financial support by
SFB 917, project A3 of the DFG, Helmholtz-Zentrum Berlin (HZB) as well as of Fonds National de la Recherche (Luxembourg).

\appendix
\section*{APPENDIX: EFFECTIVE SPIN POLARIZATION OF THE LOWER DIRAC POINT DEDUCED FROM EXPERIMENT}
\begin{figure}[tb]
\includegraphics[width=8.6cm]{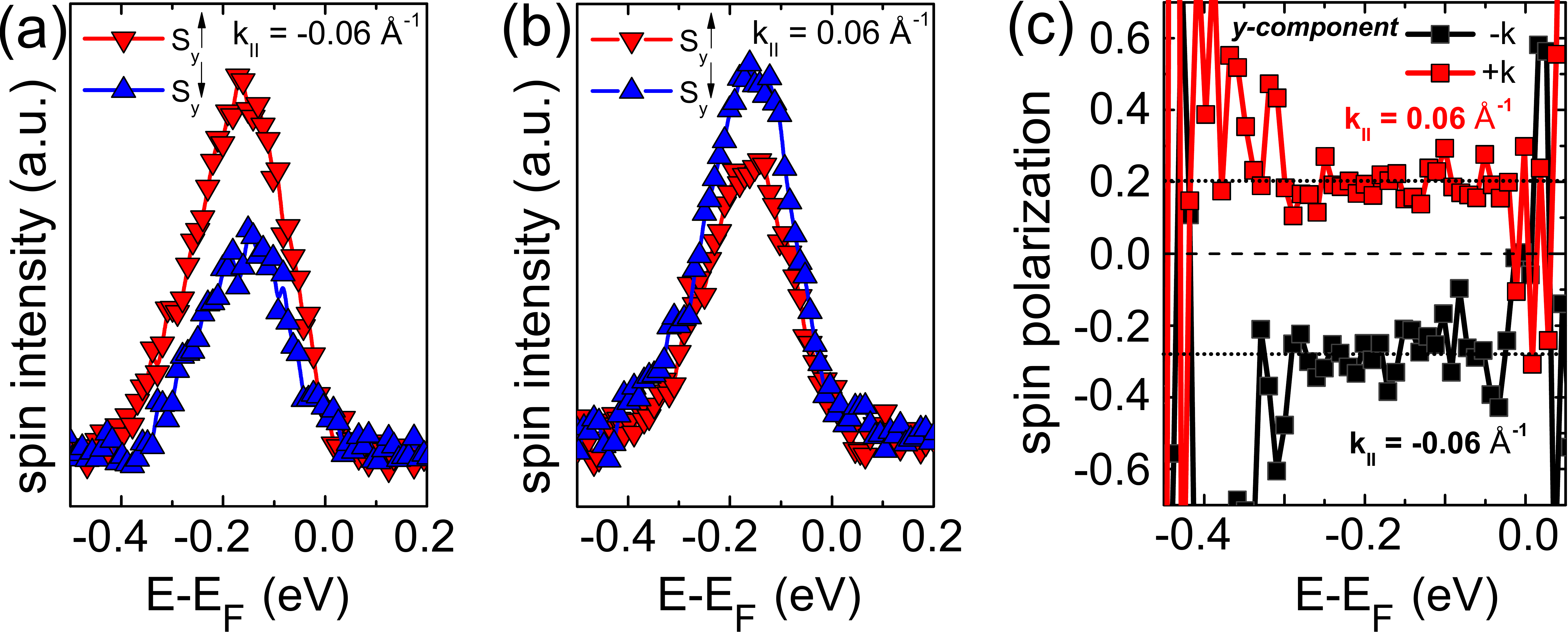}
\caption{\label{figure8}(color online) (a), (b) Spin-resolved energy distribution
curves (EDCs) for the spin component perpendicular to $k_{\rm \|}$ after subtraction of the Rashba state background (compare with Fig. 2(a) and (b)). (c) Corresponding spin polarization as a function of energy. The subtraction of the background increases the averaged spin polarization from $20$ \% to $27$ \%.}
\end{figure}

Whereas calculation predicts an in-plane spin polarization of 90\,\% for the lower part of the Dirac cone at $k_{\rm \|}=0.06$\,$\rm{\AA^{-1}}$ [Fig.~\ref{figure4}], we only detect a value of $P_y\simeq 20$ \% in the experiment. In most experimental works, the reduction of the spin polarization is due to extrinsic factors, like the insufficient instrumental resolution in spinARPES measurements especially in the case where different states are relatively close to each other \cite{Hsieh2, Hsieh3}. For surface states well separated from the bulk states, it has been demonstrated that the spin polarization reaches unity \cite{Pan}. In our case, the unpolarized background from the bulk valence band (BVB) states considerably reduces the spin polarization. Besides this background, the spin resolved spectra feature a finite background from the Rashba-type surface state visible as increasing intensity towards higher binding energies in the spin-resolved spectra in Fig. 2 (a) and (b). After subtracting this background by using a Gaussian function, the spin polarization reaches a value of $P_y\simeq 27$ \% for $k_{\rm \|}=-0.06$\,$\rm{\AA^{-1}}$ (see Fig. \ref{figure8}).

In order to discuss the absolute spin polarization more quantitatively, it is necessary to evaluate the finite contribution from the spin-degenerate BVB close to the lower Dirac cone to the ARPES intensity \cite{Souma} (Fig. 1(b) and (d)).

\begin{figure}[tb]
\includegraphics[width=8.6cm]{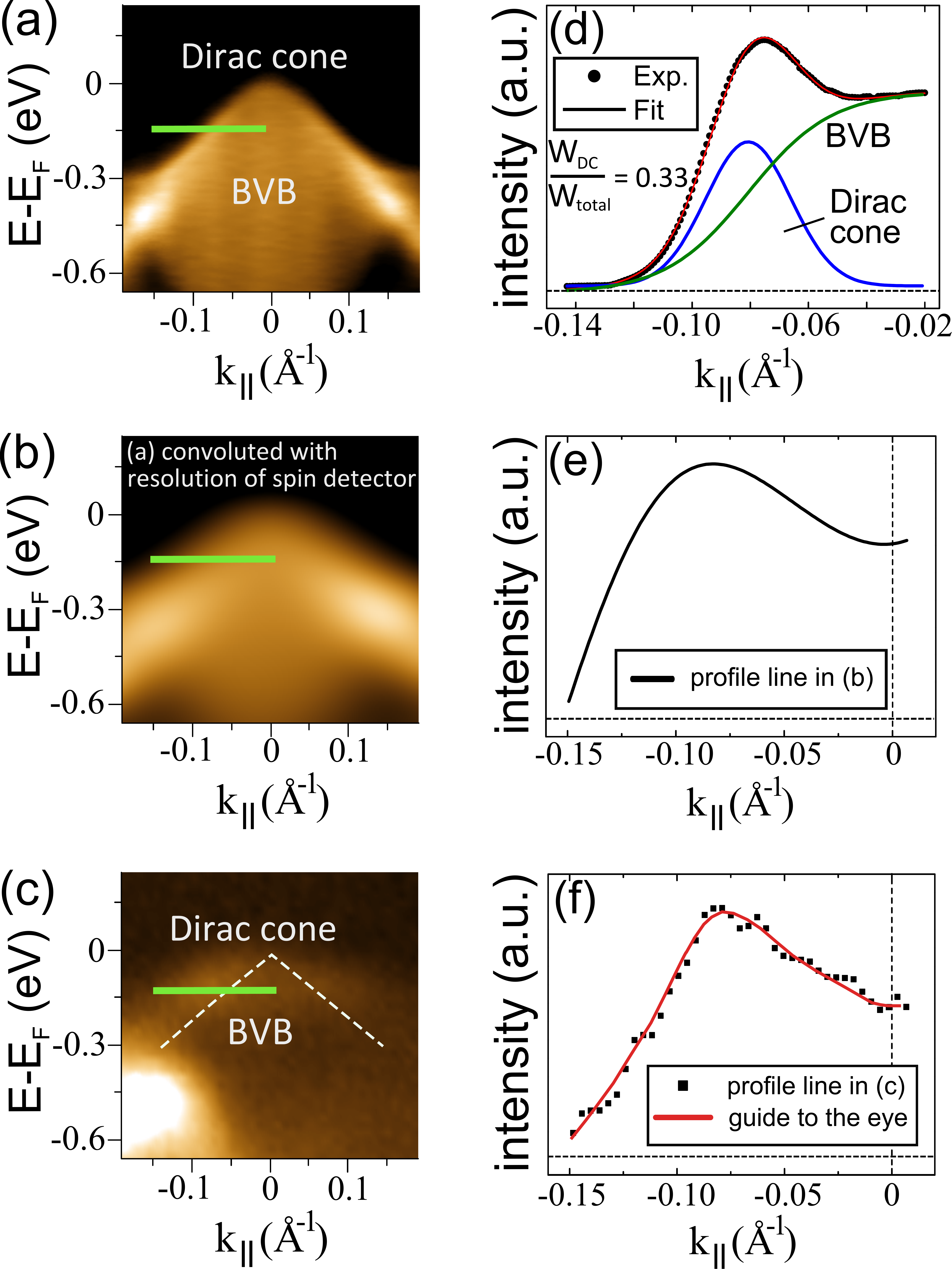}
\caption{\label{figure9}(color online) (a) High resolution ARPES data of the Dirac cone (close-up view from Fig. 1(d)) at an incident photon energy $h\nu=55$\,eV. (b) data from (a) convoluted with a Gaussian curve taking into account the energy resolution (FWHM: 100 meV) and momentum resolution (FWHM: 0.09 \AA$^{-1}$) of the spin detector. (c) ARPES data measured with the spin detector, $h\nu = 54.5$\,eV. (d)$-$(f) Constant-energy cuts through the Dirac cone along the green line in the ARPES data aside. Constant energy cuts in (e) and (f) show a similar distribution. In (d), the result of the two component fitting to the spectrum is shown as marked Dirac cone and BVB (bulk valance band).
Ratio of the spectral weight of the Dirac cone $W_{\rm{DC}}$ to the total spectral weight $W_{\rm{total}}$ is given.}
\end{figure}

Figure \ref{figure9} (a) shows a detailed ARPES measurement of the Dirac cone recorded with high resolution. The Dirac cone and the adjacent BVB are visible. If this data is convoluted with a two dimensional Gaussian function having the energy and momentum resolution of the spin resolving detector as FWHM in the two directions, the Dirac cone part overlaps much more strongly with the BVB (Fig. \ref{figure9} (b)). Indeed, a similar broadening of the structure is found in the ARPES data recorded with the spin detector (Fig. \ref{figure9} (c)), albeit with weaker intensity due to the lower efficiency of the spin resolving apparatus. Constant-energy cuts from the convoluted ARPES data (Fig. \ref{figure9}(e)) and the spinARPES data (Fig. \ref{figure9} (f)) exhibit a similar distribution, too. As a result, we feel enabled to deduce the contribution from the BVB out of the high resolution ARPES data, where we can clearly distinguish Dirac cone peak and BVB background. The quantitative analysis is shown in Figure \ref{figure9} (d) highlighting the contributions of the Dirac cone and the BVB as fitting curves using a Lorentzian for the Dirac cone states and a tanh for the BVB, respectively. In the case of the Dirac cone, the resulting peak is wider than the momentum resolution of the high resolution detector. Thus, assuming a Lorentzian function, which takes into account the lifetime broadening of the surface state, is justified. The BVB contribution is approximated by a tanh function implying a BVB of constant density within the Dirac cone as found in the DFT calculations and with no bulk bands beyond the Dirac cone. The latter implies the additional restriction within the fit that the reversal point of the tanh function matches the peak position of the Lorentzian. Width and height of the two functions are used as independent fitting parameters. From the two fit curves, shown in Fig. \ref{figure9}(d), too, we estimate the spectral weight of the Dirac cone states with respect to the total spectral weight within the spinARPES measurement. Therefore, we determine the area of each fit curve in Fig. \ref{figure9}(d) within the width of the angular resolution of the spinARPES experiment of 0.09 \AA$^{-1}$ around the probed $k_{\rm \|}$-value of 0.06 \AA$^{-1}$. They are called $W_{\rm DC}$ and $W_{\rm BVB}$, respectively. We checked that the energy resolution of the experiment is not relevant for the angular broadening. The result is $W_{\rm DC}/(W_{\rm DC}+W_{\rm BVB})=0.33$ as indicated in Fig. \ref{figure9}(d).
\begin{figure}[b]
\includegraphics[width=7cm]{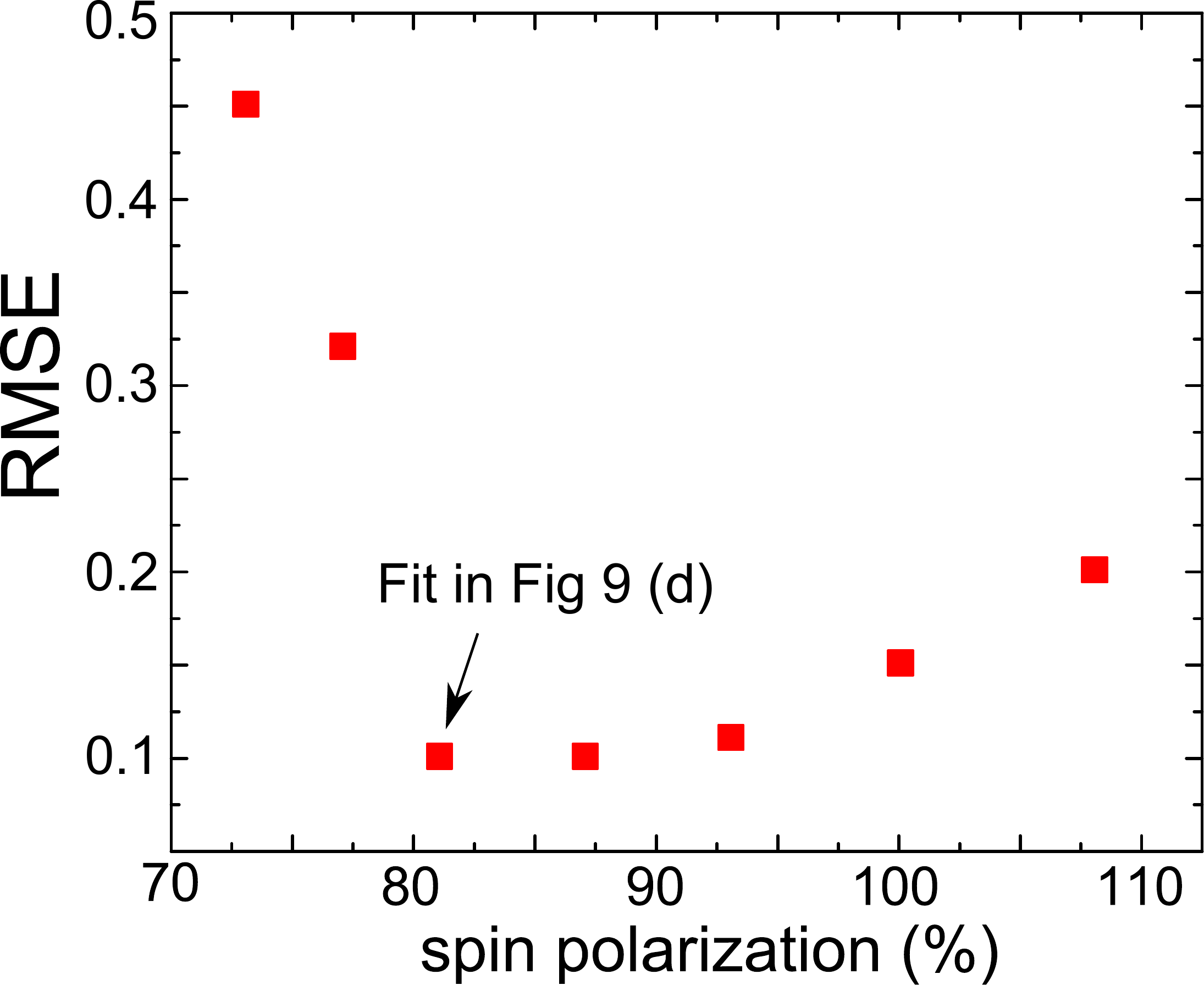}
\caption{\label{figure10}(color online) Root-mean-square error (RMSE) for different fits of the spectrum of Fig. \ref{figure9} (d)
leading to different W$_{\rm{DC}}$/W$_{\rm{total}}$ and, thus, to different spin polarizations of the Dirac cone states $P_{\rm DC}$. The RMSE increases by a factor of three for a spin polarization below 80 \%.}
\end{figure}
The resulting intrinsic spin polarization  of the Dirac cone in-plane component is then given by the experimentally measured spin polarization
$P_y$ divided by $W_{\rm{DC}}$/$W_{\rm{total}}$ leading to a value of $P_{\rm{DC}}\simeq 82$ \%. The development of the
root-mean square error (RMSE) of the fitting, as shown e.g. in Fig. \ref{figure9}(d), is then analyzed for fixed different relative heights of the Lorentzian and the tanh function, which corresponds to different $P_{\rm DC}$. The RMSE is shown as a function of $P_{\rm DC}$ in Fig. \ref{figure10} revealing that a spin polarization of 80-95 \% of the Dirac cone states is most likely. Notice the nice agreement with the DFT result of $P_{\rm DC}=90$ \%.
Notice, moreover, that the energetic width of the Dirac cone peak in spinARPES (Fig. \ref{figure8}) is similar to the total width of the BVB in the DFT calculations shown in Fig. 1(b) of the main text, which supports that our estimate of $P_{\rm DC}$ is reasonable, i.e. the BVB contribution leads to a peak of similar width as the Dirac cone peak. In turn, high resolution spinARPES experiments are required to measure the intrinsic spin polarization of the Dirac cone directly, i.e., without relying on any assumption.


\end{document}